\documentclass[a4paper,10pt,twoside]{cpc-hepnp}

\usepackage{multicol}
\usepackage{graphicx}
\usepackage{booktabs}
\usepackage{amssymb,bm,mathrsfs,bbm,amscd}
\usepackage[tbtags]{amsmath}
\usepackage{lastpage}
\usepackage{yhmath}

\begin{document}

\fancyhead[co]{\footnotesize YAN Xin-Hu~ et al:Radiation and Ionization Energy Loss Simulation for GDH Sum Rule Experiment in Hall-A at Jefferson Lab }


\title{Radiation and Ionization Energy Loss Simulation for the GDH Sum Rule Experiment in Hall-A at Jefferson Lab\thanks{Supported by the National Natural Science Foundation of China(11135002,11275083), US Department of Energy contract DE-AC05-84ER-40150 under which Jefferson Science Associates operates the Thomas Jefferson National Accelerator Facility and Natural Science Foundation of An'hui Educational Committee(KJ2012B179)
}}

\author{%
       YAN Xin-Hu(ãÆл¢)$^{1;1)}$\email{yanxinhu@ mail.ustc.edu.cn}%
\quad YE Yun-Xiu(Ò¶ÔÆÐã)$^{2}$
\quad CHEN Jian-Ping(³Â½£Æ½)$^{3}$
\quad LV Hai-Jiang(ÂÀº£½­)$^{1}$\\
\quad Zhu Pengjia(ÖìÅô¼Ñ)$^{2}$
\quad Jiang Fengjian(½¯·å½¨)$^{1}$
}
\maketitle

\address{%
$^1$Department of Physics, Huangshan University, Huangshan, Anhui 245041, China\\
$^2$Department of Modern Physics, University of Science and Technology of China, Hefei, Anhui 230026, China\\
$^3$Thomas Jefferson National Accelerator Facility, Newport News,VA 23606, USA)\\
}

\begin{abstract}
 The radiation and ionization energy loss are presented for single arm Monte Carlo simulation for the GDH sum rule experiment in Hall-A at Jefferson Lab. Radiation and ionization energy loss are discussed for $^{12}C$ elastic scattering simulation. The relative momentum ratio $\frac{\Delta p}{p}$ and $^{12}C$ elastic cross section are compared without and with radiative energy loss and a reasonable shape is obtained by the simulation. The total energy loss distribution is obtained, showing a Landau shape for $^{12}C$ elastic scattering. This simulation work will give good support for radiation correction analysis of the GDH sum rule experiment.
\end{abstract}



\begin{multicols}{2}

\section{Introduction}
 The Gerasimov-Drell-Hearn (GDH) sum rule\cite{GDH1,GDH2,Fermi} applied to nuclei relates the total cross section of circularly polarized photons on a longitudinally polarized nucleus to the anomalous magnetic moment of the nucleus:
\begin{equation}
\int_{thr}^{\infty}(\sigma_{A}(\nu,Q^{2})-\sigma_{p}(\nu,Q^{2}))\frac{d\nu}{\nu}=-4\pi^{2}\frac{\mu_{A}^{2}}{J}
\end{equation}
where $Q^{2}=-(p-p^{\prime})^2$ is the negative four-momentum squared of the exchanged photon; $p$ and $p^{\prime}$ are the four-momenta of the incoming and scattering electrons, respectively; $\sigma_{p}$ and $\sigma_{A}$ are the total photo-absorption cross sections of the nucleus with nuclear spin $J$ parallel and antiparallel, respectively, to the photon polarization; and $\mu_{A}=\mu-Jq\bar{h}/M$ is the anomalous magnetic moment of the nucleus, where q and M are the charge and mass of the nucleus. The lower limit is the photo-nuclear disintegration threshold.

  In order to obtain precise cross sections from GDH experiments, radiation correction analysis is important. In this article, we will discuss a radiation energy loss simulation based on the Single Arm Monte Carlo (SAMC) package for the GDH experiment in Hall-A at Jefferson Lab.

\section{Radiation and Ionization Energy Loss Simulation by SAMC}
SAMC is a Monte Carlo package which simulates one of the two Hall-A HRS (High Resolution Spectrometers) at Jefferson Lab. In this article, we focus on the Hadron arm (i.e. the left arm in Hall-A). SAMC works by the following procedure. Firstly, the kinematic domain illuminated and the region of interest for the analysis are defined in the input files. Secondly, the relevant variables are randomly drawn with a uniform distribution. All these variables define an event. The event undergoes different checks to see if it reaches the HRS focal plane without being stopped by the various components within the spectrometer. If it passes, the event is reconstructed at the target and stored in the output file. Meanwhile, radiation and ionization energy losses are applied each time the electron goes through some material. Before storing the event, a weight corresponding to the cross section of the event and an asymmetry can be assigned. This option is set on or off using the input file. Physics can be added into the Monte Carlo results using this weighting factor (cross section effect) or asymmetry. They are both computed for each event according to its target reconstructed kinematic quantities. Some physical procedures such as $^{12}C$ elastic cross sctions, radiative corrections and Landau tail for elastic peaks, $^{3}He$ quasi-elastic cross sections, asymmetries and external radiation corrections can be processed in this simulation package. The physics principle of the SAMC package is same as the general detector simulation toolkit, Geant4. But the SAMC is more simple, flexible and suitable for some special simulations of Hall-A at JLab. The SAMC package has included the forward and backward matrix(Optics of the HRSs) to guide the electron to go to the detector plane from the target plane for the HRS. So we can get the simulation results in the detector plane of HRS in Hall-A at JLab. That will be more useful and efficient for the data analysis. The radiation and ionization energy loss are discussed below\cite{rad0}.

\subsection{Radiation Energy Loss}
The distribution of incident electron energy loss due to bremsstrahlung in the Coulombic fields of an atom depends on frequency \cite{rad1,rad2,rad3}. The relation between the energy loss and frequency is expressed as follows.

\begin{equation}
d^{2}E=-h\nu[N]\frac{d\sigma(\nu,E)}{d\nu}dxd\nu
\end{equation}
 where $\nu$ is the frequency of radiated photons and $[N]$ is the number density of atoms. After integrating over the entire frequency spectrum of radiated photons, the energy distribution of the incident electron can be expressed as follows when it goes through the material:

\begin{equation}
{\frac{dE}{E}=-[N]\sigma_{rad}dx\rightarrow E(x)=E(0)exp(-[N]\sigma_{rad}(Z)x)}
\end{equation}
where the $\sigma_{rad}(Z)$ depends almost solely on the charge of the nucleus Z when the incident electron energies are larger than 50 MeV\cite{rad4,rad5,rad6,rad7,rad8}. We can make a assumption that the incident electron can only scatter with one atom at a time in uniform material. Under this assumption, the total energy loss can be expressed as follows:
\begin{equation}
\begin{split}
E(x)=E(0)exp(-\underset{k}{\sum}[N]_{k}\sigma_{rad}(Z_{k})x)
\end{split}
\end{equation}

where $k$ represents different types of atomic isotope. The $radiation$ $length$ of the material, which is the thickness needed for an electron to lose $1-1/e$ of its initial energy. $X_{0}$ represents the radiation length in mass per unit ``area'':

\begin{equation}
 X_{0}=\frac{A}{N_{A}\sigma_{rad}(Z)}
\end{equation}

Consequently, the unitless $radiation$ $thickness$ is defined as:

\begin{equation}
t=\frac{\rho l}{X_{0}}
\end{equation}
where $l$ is the thickness of the material.

\subsection{Ionization Energy Loss}
 When incident electrons are scattered by atomic electrons in the material, the struck atom can be ionized. The mean ionization energy loss per unit mass density per unit thickness is defined as follows\cite{rad9}:

\begin{equation}
\left[\frac{\Delta}{\rho x}\right]=\left[\frac{\xi}{\rho x}\right]\left[2log(\frac{pc}{I})-\delta(X)+g\right]
\end{equation}

\begin{equation}
\left[\frac{\xi}{\rho x}\right]=\frac{Za}{A\beta^{2}}
\end{equation}

where $\Delta$ is the mean ionization energy loss; $Z$ is atomic number; $A$ is the molecular weight of the material; $p$ is the electron's momentum; $I$ is the mean excitation potential of the material; $\delta(X)$ is the density correction \cite{rad10}; and  $\xi$ is the ``$collisional$" thickness. Since the mean energy loss is given by the Bethe-Bloch equation and the most probable energy loss is given by Landau's energy-loss formula, $g$ can be expressed in different ways\cite{rad11}. The density correction parameters from \cite{rad12} and \cite{rad13,rad14} are used. The simulation will base on the above general methods to calculate the radiation and ionization energy loss.

\section{SAMC Simulation Results and Discussion}
In order to run the SAMC simulation, we need one physical input file, ``C12.inp'' which contains the physics parameters and kinematic domain (i.e. illumination area). In this article, we only study $^{12}C$ elastic scattering before and after radiative energy loss. For convenience, the description of radiative energy loss will include radiation and ionization energy loss. The main parameters in ``C12.inp'' are shown in Table 1.

Table 1. SAMC Simulation Parameters.
\begin{tabular}{ccc}
\hline
Parameters & Value & Definition \\\hline
$N_{trail}$ & 2000000 & Number of events for\\
            &         & the kinematic domain \\
$E_{i}$     &1.14876 GeV & Beam energy \\
$E_{p}$     &1.14875 GeV/c & HRS central\\
            &          &momentum setting \\
$th_{spec}$ & $-5.99^{\circ}$ & HRS angle \\
$dpp_{ac}$  & $5\%$  & Relative momentum$\frac{\Delta p}{p}$ \\
$dth_{ac}$  & 110 mR & Vertical angle range \\
$dph_{ac}$  & 50 mR  & Horizontal angle range \\
$spot_{x}$ &0.00004 m &Total rastering size  \\
 &  &  in horizontal direction \\
$spot_{y}$ &0.00004 m &Total rastering size  \\
 &  &  in vertical direction \\
 $tgt_{l}$ &0.5 cm  & Target length  \\
 &  &  in the z direction \\
aspin & $0^{\circ}$  & Target polarization  \\
 inl& 0.00247 $g/cm^{2}$ & Thickness of matter \\
 &  & crossed by incoming $e^{-}$ \\
outl & 0.0199 $g/cm^{2}$& Thickness of matter \\
 &  & crossed by scattered $e^{-}$ \\
xdi& 0.105 $g/cm^{2}$ & Thickness of ionization  \\
& & for incoming $e^{-}$ \\
xdo& 0.798 $g/cm^{2}$ & Thickness of ionization  \\
& & for scattered $e^{-}$ \\\hline
\end{tabular}
 \\
  \\

  The beam profile is shown in Fig. 1 with a circular raster pattern. The beam size in both the x and y directions is 0.004 cm. The energy of incoming electrons is spread due to the beam energy dispersion. The external bremsstrahlung, ionization and internal bremsstrahlung are then applied. The beam energy dispersion is taken as $3 \times 10^{-5}$.
\begin{center}
\includegraphics[width=6cm,height=6cm]{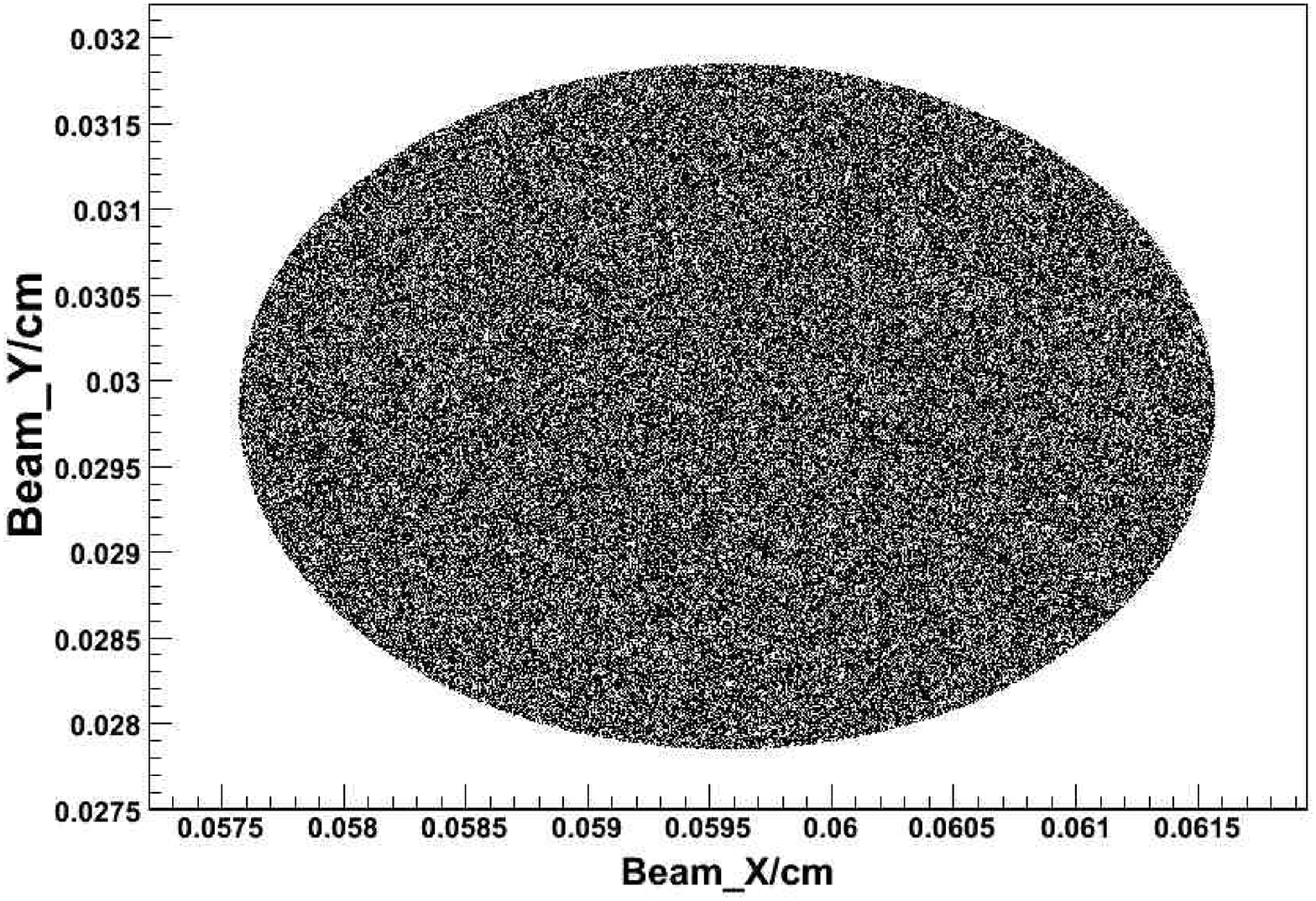}\\
Fig.~1. \quad Beam position and size distribution in simulation.
\end{center}

In this simulation, the code treats elastic scattering as different from other physics process such as quasi-elastic scattering, because of the correlation between the scattering angle and the outgoing electron momentum.
 In this case, only the scattering angle is chosen randomly. The momentum of the scattering particle is then computed according to the angle, the mass of the target $^{12}C$ and the incoming beam energy.

\begin{center}
\includegraphics[width=8cm,height=6cm]{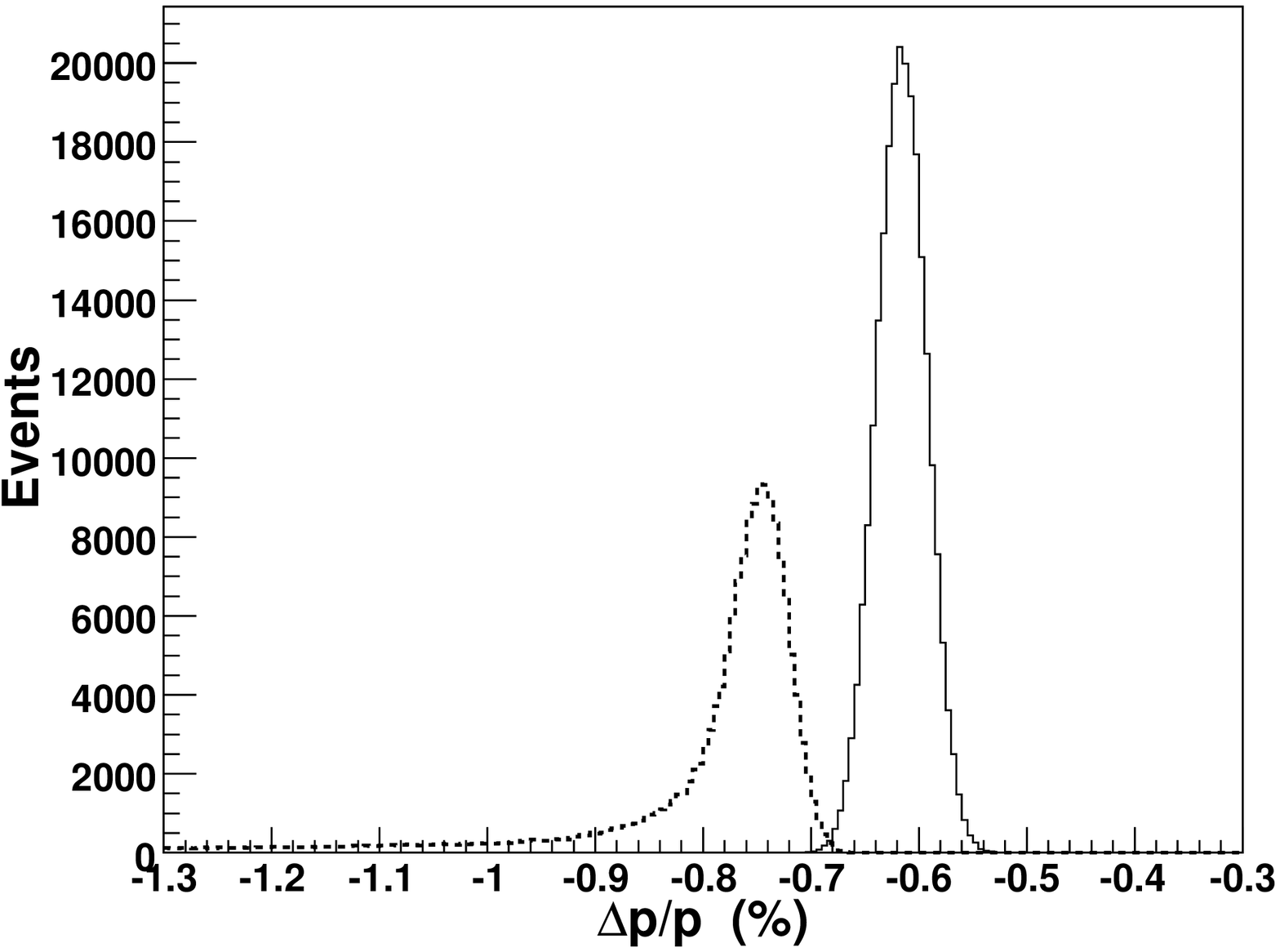}\\
Fig.~2. \quad Beam relative momemtum ratio $\frac{\Delta p}{p} (\%)$ without (solid line) and with (dashed line) radiative energy loss at target.
\end{center}

  Fig. 2 shows the beam relative momentum ratio $\frac{\Delta p}{p}$ distributions when the beam central momentum is equal to 1.14876 GeV/c. The relative momentum ratio depends on the different momentum settings, and can be tuned by the accelerator. Adjusting this momentum setting, the solid line shows the relative momentum ratio without radiative energy loss at the target, with the peak at $-0.615 \pm 0.023 \%$; the dashed line shows the relative momentum ratio with radiative energy loss at the target, with the peak at $-0.749 \pm 0.027 \% $. Obviously, we can see that the $\frac{\Delta p}{p}$ distribution gets wider after adding the radiative energy loss procedure.

\begin{center}
\includegraphics[width=8cm,height=6cm]{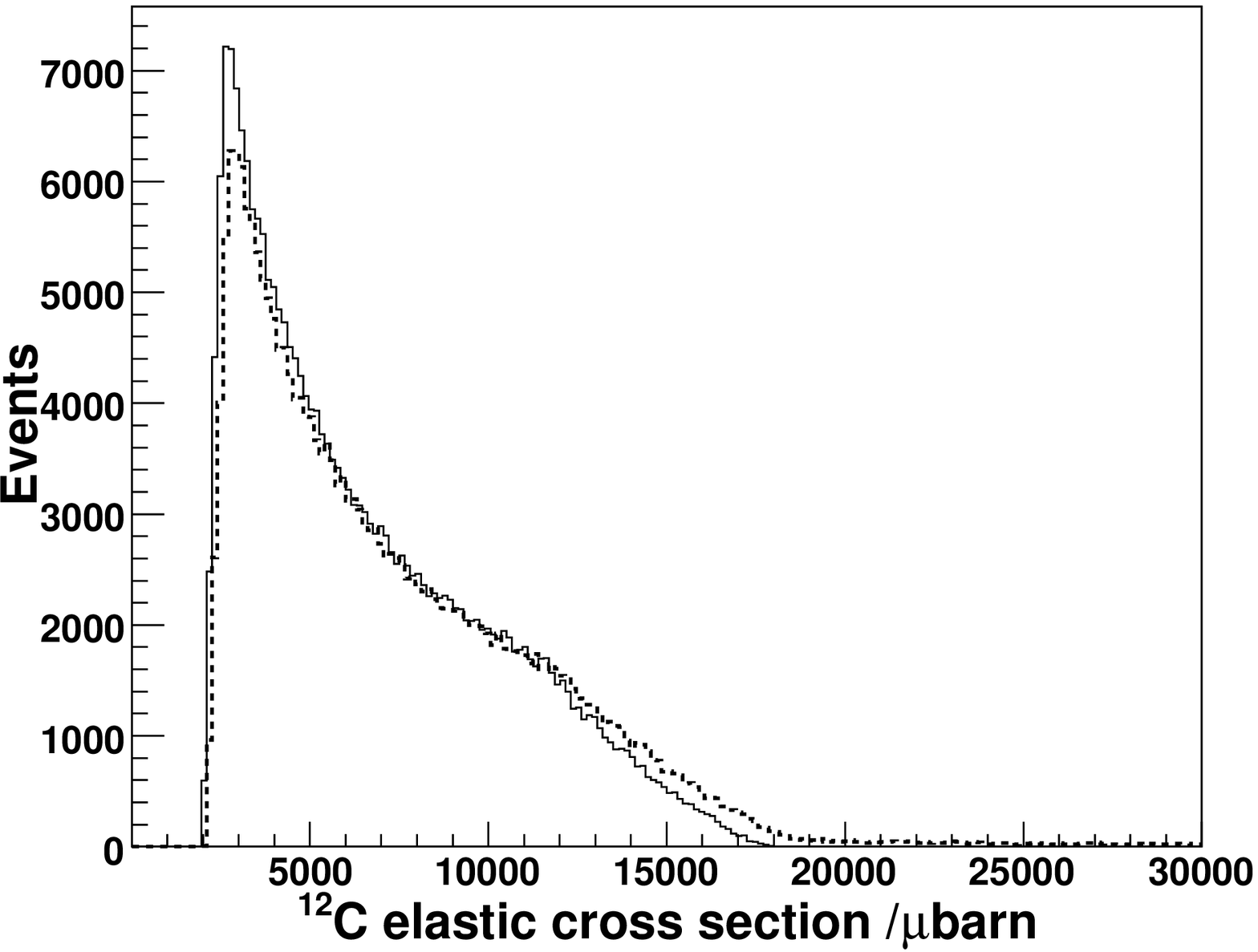}\\
Fig.~3. \quad $^{12}C$ elastic cross section without (solid line) and with (dashed line) radiative energy loss.
\end{center}

  Fig. 3 shows the $^{12}C$ elastic cross section distribution. The solid and dashed lines show the distribution without and with radiative energy loss, respectively. From the plot, we can see that the two peaks are both at around 2869 $\mu barn$. The dashed line is a little lower than the solid line, but goes a little higher than the solid line when the cross section value increases due to the radiative energy loss.

\begin{center}
\includegraphics[width=8cm,height=6cm]{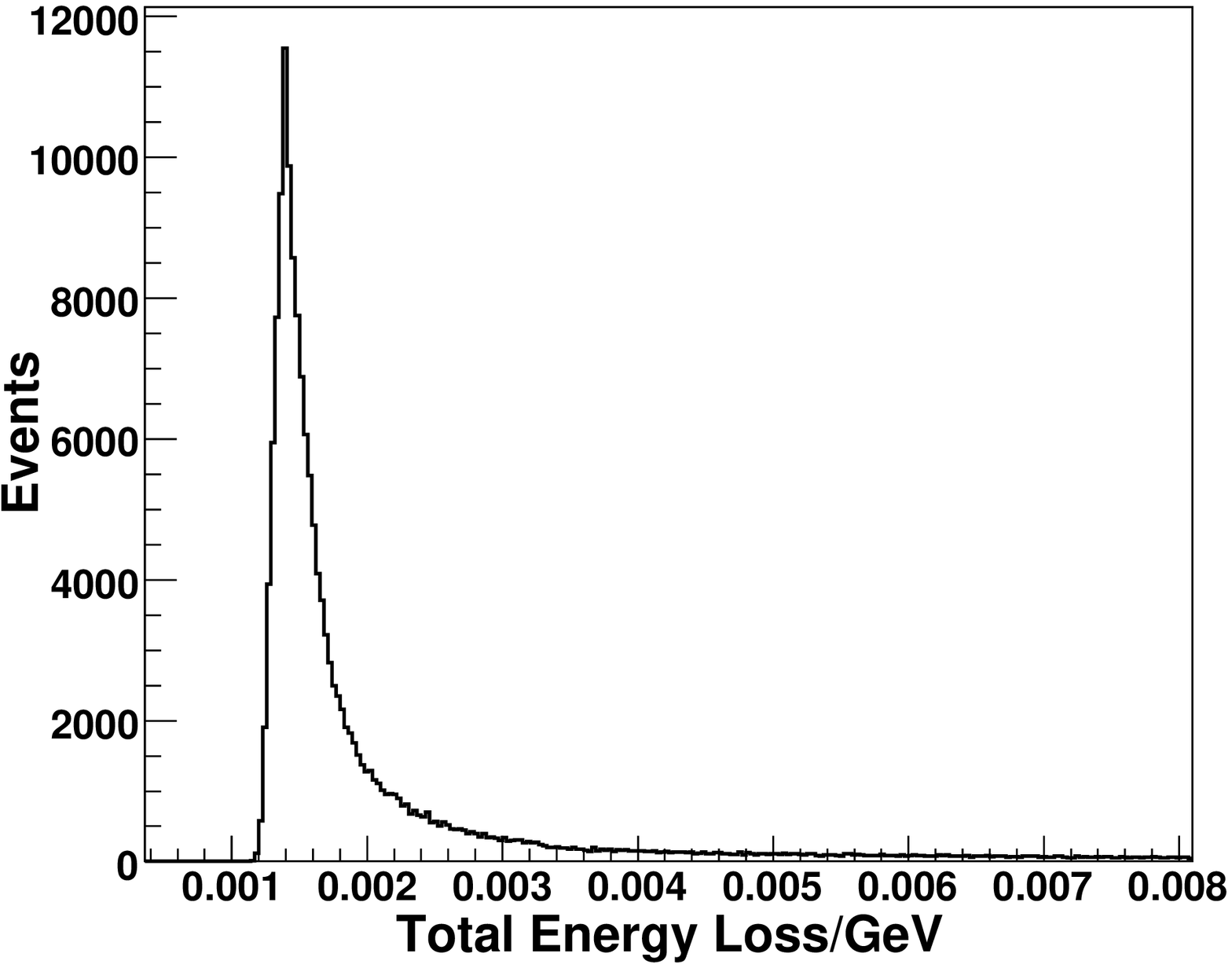}\\
Fig.~4. \quad Total energy loss distribution including radiation and ionization energy loss.
\end{center}

  Fig. 4 shows the total energy loss distribution including radiation and ionization energy loss. From the plot, we can see that the most probable total energy loss value is around $0.00142 \pm 0.00009$ GeV. The curve can be mainly fitted by a Landau distribution.

\begin{center}
\includegraphics[width=8cm,height=6cm]{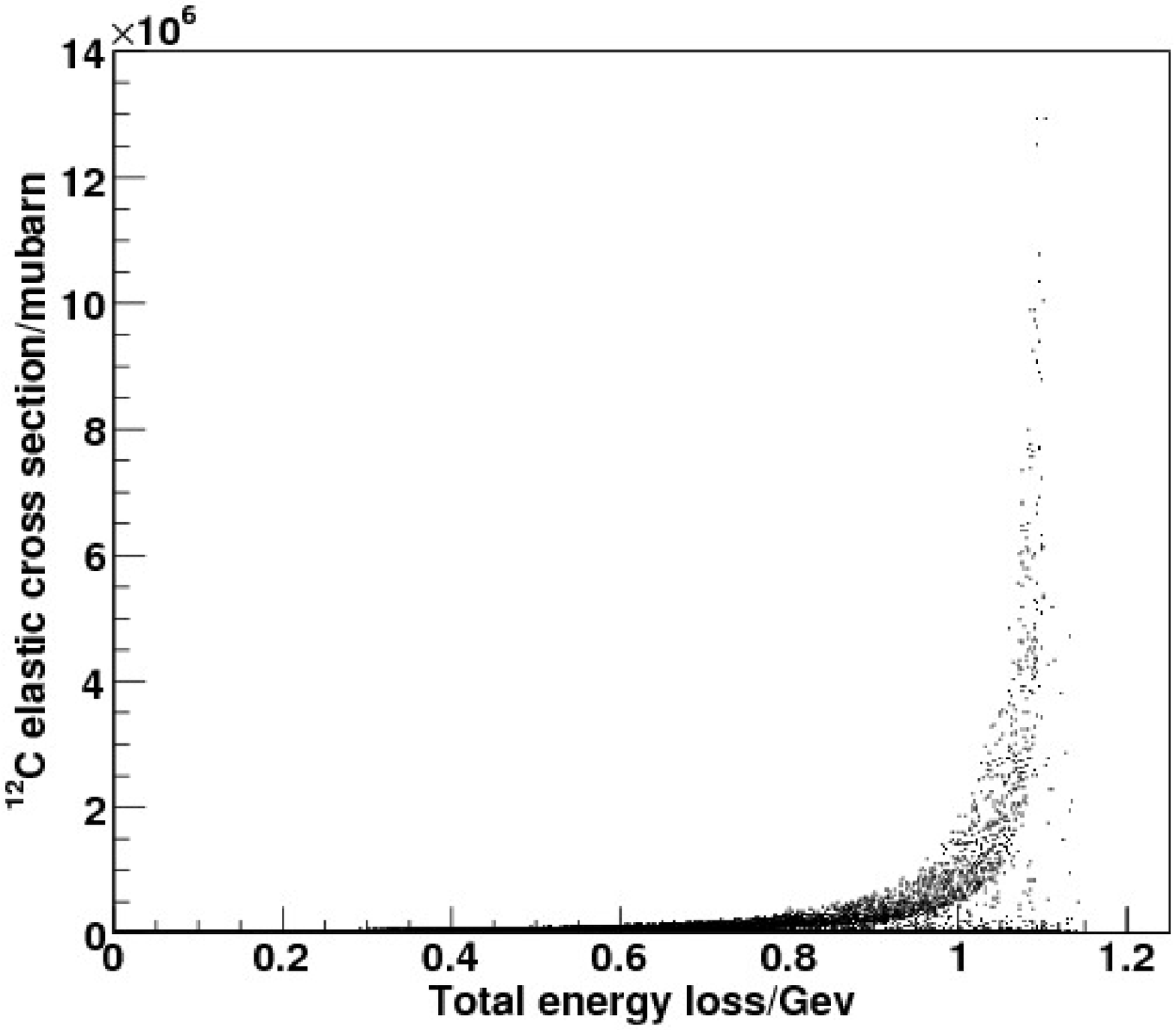}\\
Fig.~5. \quad $^{12}C$ elastic cross section vs total energy loss with radiative energy loss.
\end{center}

  Fig. 5 shows a two dimensional distribution of $^{12}C$ elastic cross section versus total energy loss after the radiative energy loss procedure. From this plot, we can see that the cross section value increases when the total energy loss of particles gets higher, but most of the events are at lower total energy loss and cross section value.

   The above results are from SAMC simulation for $^{12}C$ elastic scattering without and with radiative energy loss. We have included internal radiation (vacuum polarization, vertex corrections) energy loss and external radiation (ionization and bremsstrahlung) energy loss. These simulation results can be compared with the $^{12}C$ elastic data of the GDH sum rule experiment in future. If the simulation results match the elastic data well, this will show the good state of the data quality. The radiative corrections take the general form $\sigma_{exp} = (1+\delta)\sigma_{Born}$ where the $\delta$ represents the sum of the internal and external radiative corrections. We can make a comparison between the data and the physical theory after finishing cross section analysis. This simulation work will be helpful for the above data analysis.

\section{Summary}
This article studied the radiation and ionization energy loss based on single arm Monte Carlo simulation for the GDH sum rule experiment in Hall-A at Jefferson Lab. The radiation and ionization energy loss were discussed for $^{12}C$ elastic scattering simulation. The relative momentum ratio $\frac{\Delta p}{p}$ and $^{12}C$ elastic cross section were compared without and with radiative energy loss and a reasonable shape was obtained by the simulation. The total energy loss (including radiation and ionization) distribution was obtained, giving a reasonable distribution with a Landau shape for $^{12}C$ elastic scattering. This simulation work will provide good support for radiation correction analysis of the GDH sum rule experiment.

\acknowledgments{ We would like to thank J.P. Chen for his
    suggestions and discussion, and J. Singh and V. Sulkosky for their help.}\\

\end{multicols}

\vspace{10mm}
\vspace{-1mm}
\centerline{\rule{80mm}{0.1pt}}
\vspace{2mm}

\begin{multicols}{2}

\end{multicols}

\clearpage

\end{document}